\documentstyle[12pt,epsfig]{ioplppt_my}

\newcommand {\w} {\omega}
\newcommand {\pw} {{\omega^{\prime}}}

\newcommand {\W} {\Omega}
\newcommand {\E} {\varepsilon}

\newcommand {\ee} {{\rm e}}

\newcommand {\bfe}{{\bf e}}
\newcommand {\bfk} {{\bf k}}
\newcommand {\bfp} {{\bf p}}
\newcommand {\bfq} {{\bf q}}
\newcommand {\bfr} {{\bf r}}

\newcommand {\bfB} {{\bf B}}
\newcommand {\bfn} {{\bf n}}
\newcommand {\bfv} {{\bf v}}

\newcommand {\bsigma} {\vec{\sigma}}
\newcommand {\bxi} {\vec{\xi}}
\newcommand {\brho} {\vec{\rho}}

\newcommand {\vphi} {{\varphi}}
\newcommand {\vtheta} {{\vartheta}}

\newcommand {\Ai} {{\rm Ai}}
\newcommand {\pAi} {{\rm Ai}^{\prime}}

\begin{document}

\sloppy

\jl{2}

\title{
The influence of the spin-flip transitions on the photon spectrum 
from ultra-relativistic electrons in the field of a crystal. }

\author{
Andrei V. Korol\dag\ddag\ftnote{5}{E-mail:
korol@th.physik.uni-frankfurt.de,\ korol@rpro.ioffe.rssi.ru},
Andrey V. Solov'yov\dag\S\ftnote{6}{E-mail: 
solovyov@th.physik.uni-frankfurt.de,\  solovyov@rpro.ioffe.rssi.ru},
and Walter Greiner\dag\ftnote{7}{E-mail: greiner@th.physik.uni-frankfurt.de}
}

\address{\dag Institut f\"ur Theoretische Physik der Johann Wolfgang
Goethe-Universit\"at, 60054 Frankfurt am Main, Germany}

\address{\ddag Department of Physics,
St.Petersburg State Maritime Technical University,
Leninskii prospect 101, St. Petersburg 198262, Russia}

\address{\S A.F.Ioffe Physical-Technical Institute of the Academy
of Sciences of Russia, Polytechnicheskaya 26, St. Petersburg 194021,
Russia}

\begin{abstract}
We investigate the influence of the
spin-flip transitions on formation of the spectral distribution
of the radiation emitted by ultra-relativistic of tens to hundreds 
GeV electrons incident along  crystallographic axis in a thin 
single crystal.
Both the formalism and the numerical data are presented.
The calculated spectra for $35$ to $243$ GeV electrons in W crystal 
are compared with the dependences obtained experimentally 
\cite{MikkelsenPRL}. 
\end{abstract}

\section{Introduction}

The aim of this paper is to carry out quantitative analysis of 
the role of the 
spin-flip transitions in formation of the total emission spectra
by ultra-relativistic electrons which traverse single crystal close to its axial
direction.
This study was stimulated by a recent paper \cite{MikkelsenPRL} 
where the experimental data on radiative energy loss of  
$\E=35$ to $243$ GeV  electrons incident on a W single crystal are 
presented and, 
to a great extent, are interpreted as the first experimental evidence of a
significant contribution of spin to the radiative spectrum.

The photon emission by a highly relativistic particle in a static field of 
crystalline axis/plane and the 
magnetic bremsstrahlung (or synchrotron radiation) are examples
of the radiative processes occurring in strong fields.
It is a well-established  fact (e.g. \cite{Land4,Sokolov}) that 
quantum effects start manifesting themselves  in the photon spectrum
when a dimensionless quantum  parameter $\chi$ exceeds the value of
$\approx 0.1$. 
The quantum parameter can be written as $\chi=\gamma\, H/H_0$ 
(for magnetic bremsstrahlung) or $\chi=\gamma\, E/E_0$ (for the radiation process
in the external electric field), where 
$\gamma=\E/mc^2$ is the relativistic factor of 
projectile, $H/E$ is the strength of the magnetic/electric field and 
$H_0=E_0= m^2c^3/e\hbar$ is the critical field. 


In this paper we focus  attention on the influence of the
spin-related effects on the formation of the spectral distribution
of unpolarized photons emitted by tens to hundreds GeV 
electrons incident under small angle $\vtheta_0$ to crystallographic 
axes of a thin single crystal.

Theoretical approaches suitable for the analysis of the influence of 
the spin-related effects on the radiative spectrum of ultra-relativistic 
particles in a strong external field include (a) the quasi-classical 
method based on the operator representation of the projectile's wave 
function \cite{Baier1973,Baier} (see also \cite{Katkov2001}),
(b) the approach based on the correspondence principle and utilizing the 
equivalent photon method \cite{Lindhard91} (see also \cite{Sorensen1996}),
(c) the full quantum-electrodynamical treatment of a kinetic equation for an
electron moving in the field of a crystal axis \cite{Augustin}.


In section \ref{Formalism}  we present general formulae which allow 
to estimate quantitatively  the contribution of the spin-flip transitions 
to the total radiative spectrum.
This part of the paper, which is included to make the reading easier,
contains all the essential expressions which were used to carry out the 
numerical analysis.
They represent the quasi-classical formalism described in great detail in
\cite{Baier}. 

To say it very clearly:
(a) All the formulae are from  \cite{Baier}  except one
(equation (\ref{4.22a}) below) which could not be found there. 
This refers to the explicit expression for the differential 
intensity with spin-flip.
(b) In application to the radiation emission in crystals the formulae 
contain the standard CFA (constant field approximation) and correction 
terms (for details see section \ref{Formalism}).
(c) When carrying out the numerical analysis we used both methods, 
'pure CFA' and 'CFA + corrections'.

The numerical results are presented, discussed and compared with the 
experimental data \cite{MikkelsenPRL} in section
\ref{NumericalData}.
A summary of the results and possible directions for
further improvement of the theoretical model 
are given in section \ref{Conclusion}.
\section{The formalism\label{Formalism}}

Within the framework of the quasi-classical approach the spectral-angular
distribution of the energy radiated by an ultra-relativistic particle 
(of a spin $s=1/2$) moving in an external static field 
is given by the following formulae 
\begin{eqnarray}
{\d E\left(\bxi,\bxi^{\prime}\right) \over \hbar\d \w \,\d \W_{\bfn}}
= 
\alpha \,{\w^2 \over 4\pi^2}\,
\int \d t_1
\int \d t_2\,
R(t_2,t_1; \bxi,\bxi^{\prime})\, 
\ee^{\i\pw\,\Phi(t_2,t_1)}
\label{0.1a}\\
\bs
\fl
R(t_2,t_1; \bxi,\bxi^{\prime})
= 
{1 \over 4}\,
{\rm Sp}
\left[
\left(1+\bxi\cdot\bsigma\right)\!
\left(A^{*}(t_2)-\i\bsigma\cdot\bfB^{*}(t_2)\right)
\left(1+\bxi^{\prime}\cdot\bsigma\right)\!
\left(A(t_1)+\i\bsigma\cdot\bfB(t_1)\right)
\right]
\label{0.1b}\\
\ms
\Phi(t_2,t_1) = t_1-t_2- {1\over c}\bfn\cdot\left((\bfr(t_1)-\bfr(t_2)\right))
\label{0.1c}\\
\bs
A(t) 
= 
{c\over 2}
{\bfe^{*}\cdot\bfp(t)\over\sqrt{\E\E^{\prime}}}
\,
\left[
\left({\E^{\prime}+mc^2 \over \E+mc^2}\right)^{1/2}
+
\left({\E+mc^2 \over \E^{\prime}+mc^2}\right)^{1/2}
\right]
\label{1.8a}\\
\ms
\bfB(t) 
=
{c\over 2}
{\bfe^{*} \over \sqrt{\E\E^{\prime}}}\cdot
\left[
\left({\E^{\prime}+mc^2 \over \E+mc^2}\right)^{1/2}
\bfp(t)
-
\left({\E + mc^2 \over \E^{\prime} + mc^2}\right)^{1/2}
\left(\bfp(t)-\hbar\bfk\right)
\right]
\label{1.8b}
\end{eqnarray}
The notations used are as follows: $\alpha$ is the fine structure constant,
$\w$, $\bfk$ and $\bfe$ are, respectively, 
the photon energy, wave vector and polarization vector,
$\bfn=c\,\bfk/\w$ denotes the direction of the emission,
$c$ is the light velocity,
$\d \W_{\bfn}$ is the solid angle associated with $\bfn$.
The initial and final energies of the projectile
are denoted by
$\E$ and $\E^{\prime}=\E-\hbar\w$,
the quantities $\bxi$ and $\bxi^{\prime}$ stand for the polarization 
vectors of the initial and final states respectively, 
$\bsigma$ is the vector constructed from the Pauli matrices $\sigma_j$
($j=x,y,z$). 
The quantity $\w^{\prime}$ is defined as
$ \w^{\prime} = \w\,(\E/\E^{\prime})$ 

The remarkable feature of the quasi-classical formulae 
(\ref{0.1a}-\ref{1.8b}) is that they  combine the classical 
description of the projectile's dynamics and the specific quantum effects.
Indeed, the integrand in  (\ref{0.1a}) is time-dependent through the 
functions $\bfp(t)$ and $\bfr(t)$ which represent the classically defined
momentum and the position vector of the projectile moving in an 
external field.  
Therefore the emitted radiation is described in terms 
of the trajectory defined by $\bfp(t)$ and $\bfr(t)$, as in 
classical electrodynamics;
it is uniquely parametrized by the initial energy $\E$ and 
and the initial values 
${\bf p}_0={\bf p}(0)$ and ${\bf r}_0= {\bf r}(0)$.
 On the other hand, these expressions account for
two important types of
quantum corrections: 
(a) the radiative recoil (i.e. the change in the 
projectile  energy/momenta due to photon emission) 
which formally enters  
the integrand in  (\ref{0.1a}) through the 'corrected' photon frequency 
$\w^{\prime}$,
and (b) the dependence of the intensity of radiation on the 
spin quantum numbers of the projectile. The latter enter the formulae
through the two-component polarization density matrices
$(1+\bxi\cdot\bsigma)/2$  and $(1+\bxi^{\prime}\cdot\bsigma)/2$
(see (\ref{0.1b})).

The quantum corrections related to the projectile's dynamics in the 
external field are neglected. 
This is fully justified provided the particle
can be considered as an ultra-relativistic in both the initial and
the final states. 
Hence, it is implied that the relativistic factors satisfy the
strong inequalities
$\gamma = \E/mc^2 \gg 1$, $\gamma^{\prime} = \E^{\prime}/mc^2 \gg 1$.
If these conditions are met then, as it is proved in \cite{Baier1973,Baier}
(see also the recent paper \cite{Katkov2001}),
the expressions (\ref{0.1a}-\ref{1.8b}) can be used to 
evaluate the characteristics of radiation emitted by a projectile 
moving in an arbitrary static field.

Once the functions $\bfp(t)$ and $\bfr(t)$ are known, then  the 
these formulae allow one to evaluate the most detailed
characteristics of the emitted radiation. 
These include the dependence of $\d E$ not only on the photon
energy and the emission angle but the polarizational properties as well.
In particular, one can analyze the role of the spin degree of freedom
in the radiative process.

To obtain expressions describing the 
radiative spectra 
of a collimated bunch of unpolarized electrons passing through a crystal
close to the axial direction one has to carry out the following 
transformation of the right-hand side of (\ref{0.1a}):
(a) summation over the photon polarizations,
(b) averaging over the electron polarizations (the cases
$\xi^{\prime}=\pm\xi$ have to be considered separately),
(c) integration over the emission angles,
and (d) averaging over the fast oscillatory transverse motion 
of the electrons due to the action of the axial field.
The transformations (a)-(c) are carried out using standard methods 
(see, e.g., \cite{Land4,Baier}). 
An adequate approach for averaging over the transverse motion 
is based on the statistical description of the particle distribution 
in the phase space (\cite{Baier,Lindhard1965}). 
In a thin crystal, where one can neglect the influence of multiple 
scattering from the crystal nuclei and electrons,
the differential probability, $\d w$, for a projectile 
with (transverse) coordinate $\brho$ and velocity $\bfv_{\perp}$ 
can be written as follows
\begin{eqnarray}
\d w\left(\brho,\bfv_{\perp}\right)
= {\d^2\brho \over s}\,
 F\left(\brho,\vtheta_0\right),
\label{4.47}
\end{eqnarray}
where $s$ is the area corresponding to one axis, 
and  $\vtheta_0$ is the incident angle with respect to the axis
(see the illustrative figure 1).' 
\begin{figure}
\vspace*{0.cm}
\begin{center}
\epsfig{file=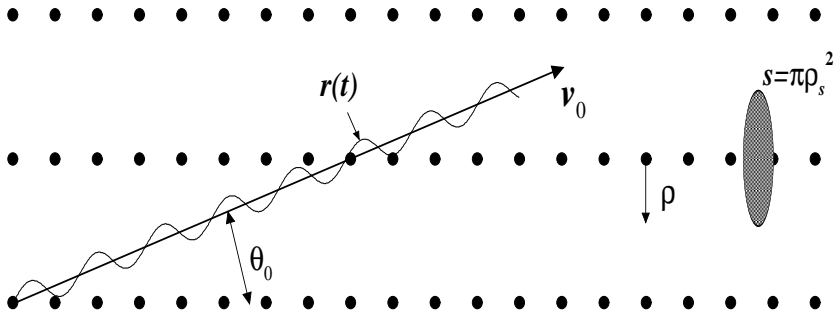, width=11cm, height=4cm, angle=0}
\end{center}
%
\caption{Schematic representation of the passage of 
an ultra-relativistic particle through an oriented 
crystal.
Under the action of the crystalline field $U$ (the filled circles
denote the positions of atomic nuclei arranged into the 
crystallographic axes, \ - \ atomic strings) 
the trajectory ${\bf r}(t)$ of the particle
varies along the direction of the vector ${\bf v}_0$,
 which is the mean velocity. 
The picture corresponds to the case when incident angle 
$\vartheta_0$ exceeds the Lindhard's angle $\vartheta_L$.
The dashed ellipse denotes the area, $s=\pi \rho_s^2$,
allocated for one axis.
The field of each atomic string 
is supposed to be axially symmetric within $s$: $U=U(\rho)$
where $\rho$ is the distance from the axis. 
}
\label{add.fig}
\end{figure}
 The distribution function
$F\left(\brho,\vtheta_0\right)$ is defined by the initial 
conditions at the entrance. 
In the case of an axially symmetric potential
$U(\rho)$ of the atomic string the function 
$F\left(\brho,\vtheta_0\right) \equiv
F\left(\rho,\vtheta_0\right)$ is given by:
\begin{eqnarray}
 F\left(\rho,\vtheta_0\right)
=
\int {\d^2 \brho_0 \over s\left(\E_{\perp 0}\right)}\,
\theta\left(\E_{\perp 0}-U(\rho)\right),
\label{4.48}
\end{eqnarray}
where 
$s\left(\E_{\perp 0}\right)=
\int \d^2 \brho\ \theta\left(\E_{\perp 0}-U(\rho)\right)$
 is the allowed area for the transverse motion in a unit cell,
$\theta(x)$ is the Heaviside function:
$\theta(x) = 1$ when $x>0$, and $\theta(x) = 0$ if otherwise.
The quantity $\E_{\perp 0}$ stands for the transverse energy 
$\E_{\perp\, 0} = \E \vtheta_0^2/2 + U(\rho_0)$ which, for given
$\vtheta_0$ and the entrance coordinate $\rho_0$, is the constant of motion
in a thin crystal. 
Note that for the incident angles larger than the critical Lindhard-angle,
$\vtheta_L = (2U_0/\E)^{1/2}$ (where $U_0$ is the depth of the potential 
well), when $\E_{\perp 0}>U_0$, the right-hand side of 
(\ref{4.48}) reduces to unity.

Carrying out all the operations mentioned above one obtains the 
spectral distributions of the emitted energy (per path $\d l = c\, \d t$)
in the following form:
\begin{eqnarray}
\left\langle\overline{{\d E^{({\rm t})}\over\hbar\d \w\,\d l}}\right\rangle
=
{1 \over \pi \rho_s^2}\,
\int_0^{2\pi}\d \vphi_{\brho} \int_0^{\rho_s}
\rho\,\d\rho\,
 F\left(\rho,\vtheta_0\right)
\overline{{\d E^{({\rm t})} \over \hbar\, \d \w\, \d l } },
\label{4.22}
\end{eqnarray}
where the brackets designate the averaging over the phase space, and
the bars indicate that the averaging over the initial polarizations of the 
projectile is carried out. 
The integration 
is carried out over the area allocated for one axis, $s = \pi\rho^2_s$,
where $\rho_s$ is  half the distance between two neighbouring
axes of the same type.

The superscript '(t)'  
specifies the type of the spectral distribution: 
$({\rm t})=(\pm)$  correspond to the spectra with, (-), and without, (+),
the spin-flip transitions, 
and $({\rm t})=({\rm tot})$ designates the total spectrum
which is the sum of the (+)- and (-)-terms.

The spectral distribution appearing in the integrand of (\ref{4.22}) reads
\begin{eqnarray}
\overline{{\d E^{({\rm t})} \over \hbar\, \d \w \, \d l} }
= 
{\alpha  \over \lambda_c \gamma}\,
C^{({\rm t})}\,
{\i \over 2\pi}
\int_{-\infty}^{\infty}
{\d \tau \over \tau+ \i 0 }\, \ee^{\i\, \Phi(\tau)}
\left(1+\beta^{({\rm t})} A(\tau)\right),
\label{4.22a}
\end{eqnarray}
where 
$\lambda_c$ is the Compton wavelength of the electron.
For the cases '(t)=(tot)' and '(t)=(-)' the coefficients 
$C^{({\rm t})}$ and $\beta^{({\rm t})}$ are given in terms of
the fractional photon energy $\mu=\hbar\w/\E$, $0\leq \mu \leq 1$:
\begin{eqnarray}
\fl
C^{({\rm tot)}}= -\mu,
\quad
C^{(-)}= {\mu^3\over 6(1-\mu)},
\qquad
\beta^{({\rm tot})}=1-\mu + {1 \over 1-\mu},
\quad
\beta^{(-)}=-2.
\label{4.22d}
\end{eqnarray}
For small incident angles $\vartheta_0$ one can use the continuous
model for the potential of the axis.
Then, assuming this potential to be axially symmetric,
solving (approximately) the equation of motion 
$\dot{\bf v}(t)= \E^{-1}\,\left(\d U(\rho)/\d \vec{\rho}\right)$
using on the right-hand side the relation 
${\bf r}(t)={\bf r}_0+{\bf v}_0\, t$ for the position vector
(${\bf v}_0$ is the the mean velocity of the particle),
and, finally, using the obtained expressions in eqs. (1)-(5), 
one finds that the functions 
$\Phi(\tau)$ and  $A(\tau)$ can be represented
as follows
\begin{eqnarray}
\fl
\Phi(\tau) 
&=
{\pw \tau \over\gamma^2}
\left[
1
+
{\vtheta_v^2 \over \pi^2}
 \int \d^2 \bfq\,\d^2\bfq^{\prime}\, 
{\bfq \cdot\bfq^{\prime}\over q_0q_0^{\prime}}\,
\ee^{\i(\bfq+\bfq^{\prime})\brho}\,
 \ee^{-{ u_t^2 \over 2}(q^2+q^{\prime^2})}\,W(q)\,W(q^{\prime})\,
\right.
\nonumber\\
\fl
&\qquad
\left.
\times
\left(
{\sin\left(q_0+q_0^{\prime}\right)\tau \over 
\left(q_0+q_0^{\prime}\right)\tau }
-
{\sin (q_0\, \tau) \over q_0\, \tau}
{\sin (q_0^{\prime}\, \tau) \over q_0^{\prime}\, \tau}
\right),
\right]
\label{A.4}\\
\fl
A(\tau)
&=
- 
 {\vtheta_v^2 \over \pi^2}
 \int \d^2 \bfq\, \d^2 \bfq^{\prime}\, 
{\bfq\cdot\bfq^{\prime} \over q_0q_0^{\prime}}\,
\ee^{\i(\bfq+\bfq^{\prime})\brho}\,
 \ee^{-{ u_t^2 \over 2}(q^2+q^{\prime^2})}\,W(q)\,W(q^{\prime})\,
\sin(q_0\tau)\,
\sin(q_0^{\prime}\tau),
\label{Delta.2}
\end{eqnarray}
where $\vtheta_v=U_0/mc^2$,
$W(q)$ is the atomic formfactor,
$u_t$ is the amplitude of thermal vibrations of the atoms, and
$q_0=\bfq\cdot\bfv_0$.
The integration is carried out over the  vectors $\bfq$ and 
$\bfq^{\prime}$ lying  in the plane perpendicular to the axis.
The details of derivation of eq. (\ref{4.22a}) in the case of the 
total spectrum can be found in \cite{Baier}.

Let us note here that the radiative transitions accompanied by the 
spin-flip process manifest themselves for large energies of 
the emitted photon, $\hbar\w \sim \E$. 
Indeed, it follows from (\ref{4.22a}-\ref{4.22d}) that 
$\overline{\d E^{(-)}}\propto {\mu^3/(1-\mu)}$. 
The latter quantity is negligibly small for $\mu \ll 1$ and increases
as $\mu \sim 1$.
It is known (see, e.g. \cite{Land4,Baier}) that the emission of 
such energetic quanta becomes noticeable in the radiative spectrum 
when the quantum parameter $\chi_s$ becomes large, $\chi_s > 1$.
Hence, the magnitude of $\chi_s$ defines the effective range of the 
emitted photon energies. 
In the case of an ultra-relativistic particle passing through 
a crystal close to the direction of a crystallographic axis 
the parameter $\chi_s$ is conveniently represented in the following form:
\begin{eqnarray}
\chi_s = \gamma\,{\lambda_c\over a_s}\, {U_0 \over mc^2}.
\label{chi_s}
\end{eqnarray}
The quantity $a_s$ represents the effective screening radius of the 
potential created by a string of the crystal atoms.
For the potential of (111) axis in a tungsten crystal, where 
$U_0 \approx 400$ eV, $a_s\approx 0.2$ \AA (see, e.g. \cite{Baier}),
it follows that $\chi_s=1$ is achieved for the incident electron 
energy  $35$ GeV. 
Hence, one can expect that the effects due to
the spin-flip transitions will manifests themselves in the total radiative
spectrum of an $\E > 35$ GeV electron moving at  small angle $\vtheta_0$ 
with  respect to (111) direction in a W crystal.

>From the computational viewpoint the direct calculation of the 
functions (\ref{A.4}-\ref{Delta.2}) and, correspondingly, 
the  spectra (\ref{4.22a}) is a non-trivial problem even if one utilizes
simple analytic forms for the axial potential such as the Moli\'ere 
approximation or the Lindhard potential \cite{Lindhard1965}.
Considerable simplification of the formulae is achieved 
within a so-called constant field approximation (CFA) \cite{Baier}.
The latter implies that the change in the particle's radial distance,
$\Delta \rho = v_{\perp}\tau_c= c\vtheta_0\tau_c$, 
occurring during the time interval $\tau_c = l_c/c$ (here
$l_c \sim c\gamma^2 /\w^{\prime}$ is a coherence length), is smaller than
the typical scale $a_s$ within which the axial field changes noticeably.
Hence, provided the condition $l_c \vtheta_0 < a_s$ is met, one can expand
the integrands in  (\ref{A.4}-\ref{Delta.2}) in powers of 
$q_0\, \tau \sim c \vtheta_0 \tau/a_s <1$. 
If, in addition, the incident angle $\vtheta_0$ is small compared with 
$\vtheta_v$, then the terms proportional to 
$(\vtheta_0/\vtheta_v)^2$ can be treated as small corrections.
Within this approximation the functions (\ref{A.4}-\ref{Delta.2})
are expressed in terms of local characteristics of the trajectory.
The final result, valid for both the total spectrum and for its part
due to the spin-flip, reads 
\begin{eqnarray}
\fl
\overline{{\d E^{({\rm t})} \over \hbar\,\d \w \, \d l}}
&=
{\alpha C^{({\rm t})}\over \lambda_c \gamma}
\left\{
\int_{z}^{\infty}\!\! \Ai(\xi)\,\d \xi
+
\beta^{({\rm t})}{\pAi(z) \over z}
-
{\vtheta_0^2 \over \vtheta_v^2}
\left[
\beta^{({\rm t})} 
\biggl(a F_1(z)
-
b F_2(z)\biggr)
+
b F_3(z)
\right]
\right\}
\label{4.56a}
\end{eqnarray}
where $\Ai(.)$ and $\pAi(.)$ are the Airy function and its derivative
respectively
(see, e.g., \cite{Abramowitz}).
Other variables and functions are defined as follows:
\begin{eqnarray}
z=\left({u\over \chi(\rho)}\right)^{2/3},
\quad
u= {\mu \over 1- \mu},
\label{4.61a}\\
\chi(\rho) = \chi_s\,w^{\prime}(y),
\label{4.61b}\\
y = {\rho \over a_s},
\quad
w(y)=
{ U(\rho) \over U_0},
\label{4.61aa}
\end{eqnarray}
\begin{eqnarray}
a 
= 
 {1\over 3} {2h_1 + h_2 \over y^4 g^2},
\qquad
b 
= 
 {1\over 15}{2h_1 + h_2 + h_3\over y^4 g^2},
\label{4.61c}\\
\bs
h_1 
=
{y g^{\prime} \over g},
\quad
h_2 
=
y\,{w^{\prime\prime\prime}-3g^{\prime} \over 2g},
\quad
h_3 = {1 + (1+ h_1)^2 \over 6},
\quad
g = {w^{\prime} \over y},
\label{4.61d}
\end{eqnarray}
\begin{eqnarray}
F_1(z) 
&=
{ z\pAi(z) +\Ai(z)\over z^2},
\qquad
F_2(z)
=
{ 6z\pAi(z) +(4+ z^3)\Ai(z)\over z^2},
\nonumber\\
F_3(z) 
&=
{2z\pAi(z)+z^3 \Ai(z)\over z^2}\ .
\label{4.60}
\end{eqnarray}

By expressing 
$\Ai(z)$ and $\pAi(z)$ in terms of the modified Bessel functions
$K_{1/3}(x)$ and $K_{2/3}(x)$ with $x = 2z^{3/2}/3$ 
(see, e.g., \cite{Abramowitz}) we reproduce the
formula presented in \S 17 of \cite{Baier}.

The first two terms on the right-hand side of (\ref{4.56a}) 
are independent of the incident angle $\theta_0$
and represent the quasi-classical
spectral distribution of the radiated energy in the magnetic bremsstrahlung
limit (see, e.g. \cite{Land4,Baier}). 

The last three terms on the right-hand side of  (\ref{4.56a}) represent
the correction to the CFA due to the change in the field strength
on the scale of the coherence length, $l_c$. 
These terms appear as a result of the transformation of a more general
expression (\ref{4.22a}). 
The range of validity of (\ref{4.56a}) is subject to two conditions:
(a) $\vtheta_0 \ll \vtheta_v$, and (b) $l_c \vtheta_0 < a_s$.

\section{Numerical results \label{NumericalData}}

In this section we present the results of calculations 
of the spectral intensity of radiation emitted by ultra-relativistic
electrons in the energy range $\E=35 - 243$ GeV incident with 
$\vtheta_0 = 0.3$ mrad from (111) axis in a W crystal of a thickness
$0.2$ mm. 
The parameters correspond to those used in a recent experiment 
\cite{MikkelsenPRL} where the main emphasis was put on the investigation
of the role of the spin-flip process in forming the total radiative 
spectrum.

The calculation of the spectral intensity was based on 
the formulae (\ref{4.56a}-\ref{4.60}). 
The averaging over the projectile trajectories was carried out according
to the rule (\ref{4.22}) where we used $F\left(\rho,\vtheta_0\right)=1$
which is justified by the relation
$\vtheta_0 > \vtheta_L$. 
Indeed, for the potential of a (111) axis the depth of the potential well 
is $\approx 1000$ eV and, therefore, the 
Lindhard critical angle $\vtheta_L=0.09\dots 0.2$ mrad for 
$\E=35 - 243$ GeV. 

\begin{figure}
\vspace*{0.cm}
\begin{center}
\epsfig{file=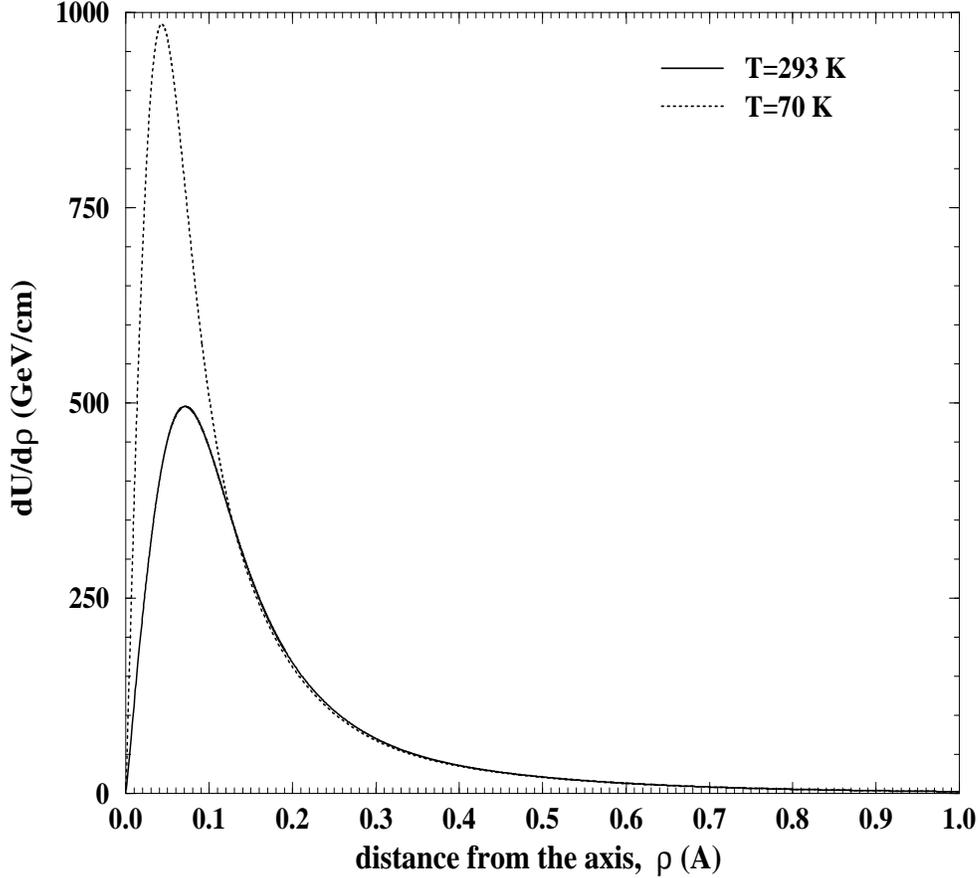, width=13cm, height=12cm, angle=0}
\end{center}%
\caption{The field strength, $\d U(\rho)/\d \rho$,
of a (111) axis in W as a function of the distance from
the axis for two different temperatures as indicated.
}
\label{dU.fig}
\end{figure}

We used a continuous potential approximation for modelling 
the field of an atomic string. 
The potential of an individual atom was calculated by using the
radial wavefunctions within the  non-relativistic  Hartree-Fock
approximation. 
For the distances $\rho=[0,\rho_s]$ (where $\rho_s=L_c/\sqrt{6}$ is the 
half-distance between two (111) axes, $L_c=3.165$ \AA\, 
is the lattice constant of a W crystal \cite{Gemmell})
the potential of the individual string was corrected by accounting for
the potentials created by the nearest six axes. 
Only axially-symmetric part of the resulting field was used.
The dependence of the field strength $\d U(\rho)/\d \rho$ on
the distance $\rho$ from the axis are plotted in figure \ref{dU.fig}.
The screening distances, $a_s$, defined as the root of the equation
$\d U(a_s)/\d \rho = e^{-1}\,(\d U/\d \rho)_{max}$, are equal to 
$0.125$ \AA\ for the crystal temperature $T=70$ K 
and to $0.191$ \AA\  for $T=293$ K. 

As functions of $\rho$ the spectral intensities 
$\overline{{\d E^{({\rm t})}/ \hbar\,\d \w \, \d l}}$ depend strongly
on the magnitude of the local quantum parameter $\chi(\rho)$ 
defined by eq. (\ref{4.61b}). 
Namely, the value of $\chi(\rho)$ defines the effective range of the emitted
photon energies. 
The formal analysis of this statement is as follows.
The right-hand side of (\ref{4.56a}) is expressed in terms of the Airy 
function and its derivative. 
These functions satisfy the conditions
 ${\rm Ai}(z),|{\rm Ai}^{\prime}(z)|=0.1\dots 1$ for  $ z\le 1$, 
and decrease rapidly (exponentially) as $z$ goes beyond 1.
The same is valid for the integral term, 
$\int_{z}^{\infty}\!\! \Ai(\xi)\,\d \xi$ (see, e.g.,\cite{Abramowitz}).
Hence, in the $\rho$-range where $\chi(\rho) \ll 1$, the spectral 
distributions increases sharply within the interval of low energies
of the emitted photon,
$\hbar\w=0\dots \hbar\w_{max}$ with 
$\hbar\w_{max}\approx \chi(\rho)\E \ll \E$, 
and decrease rapidly as $\hbar\w/\E > \chi(\rho)$.
If $\chi(\rho) \gg 1$ then the condition $z \sim 1$ produces 
(see (\ref{4.61a})) $\hbar\w_{max}/\E \sim 1$ which makes possible the 
emission of energetic quanta. In this case the total 
spectral distribution 
$\overline{{\d E^{({\rm tot})}/\hbar\,\d \w \, \d l}}$ is
(roughly) flat within the interval $\hbar\w/\E = 0\dots 1$ \cite{Baier}.
The contribution of the spin-flip term becomes noticeable
(and comparable in magnitude with the total spectrum) at the 
hard end of the spectrum  whereas  for $\hbar\w/\E \ll 1$ it is negligibly
small. The latter result immediately follows if one compares the 
coefficients $C^{({\rm tot)}}$ and  $C^{(-)}$ (see (\ref{4.22d})):
for $\mu\ll 1$ the following strong inequality is valid
$C^{(-)}/C^{({\rm tot)}}\approx \mu^2 \ll 1$.

\begin{figure}
\vspace*{0.cm}
\begin{center}
\epsfig{file=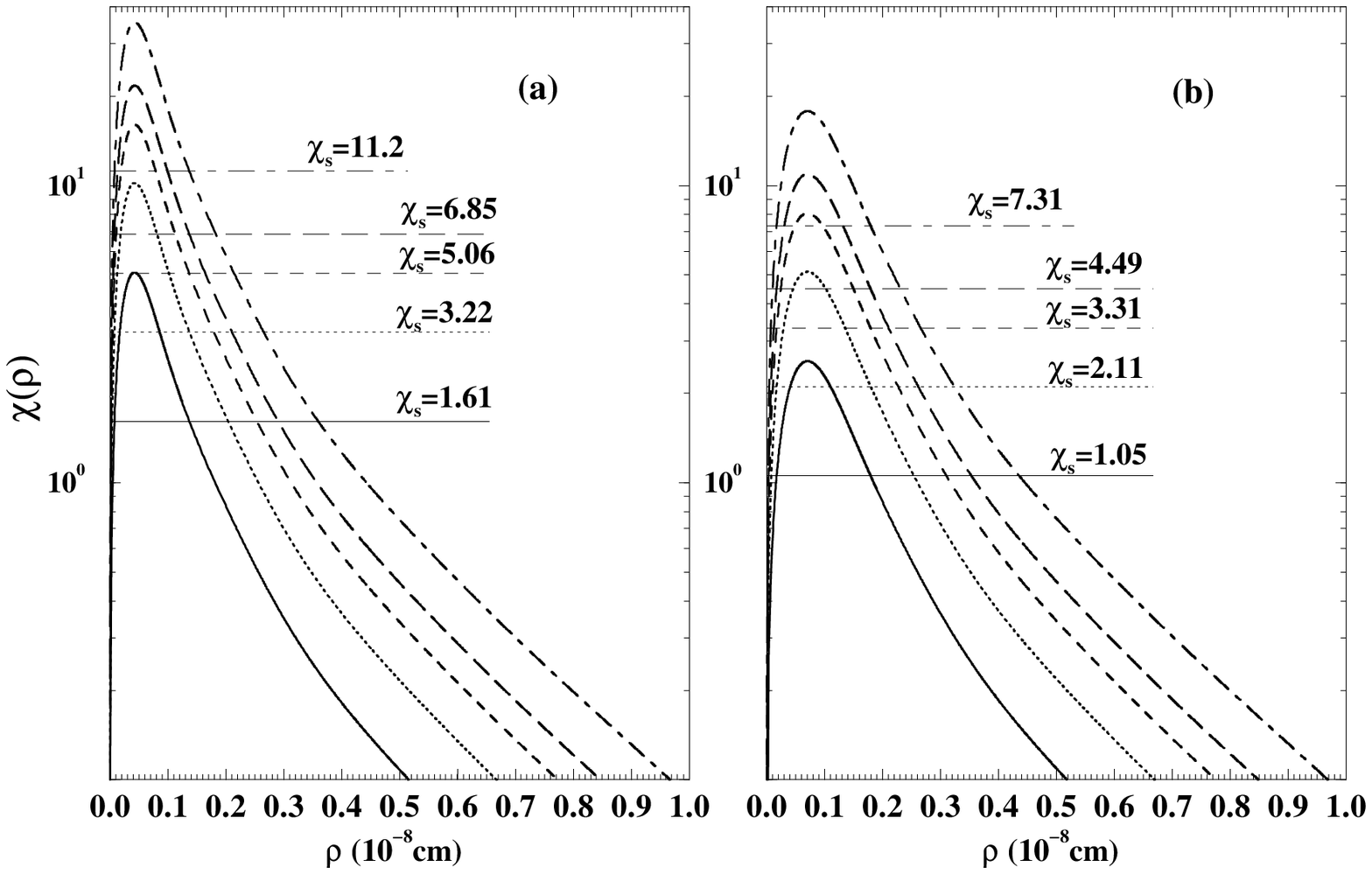, height=9cm, angle=0}
\end{center}%
\caption{The local quantum parameter $\chi(\rho)$ (see (\protect\ref{4.61b}))
versus  the distance from the axis $\rho$ for the field of
(111) axis in W calculated for $T=70$ (figure (a)) and for 
$T=293$ K  (figure (b)), and for
different incident energies: 
$\E=35$ GeV (solid lines),
$\E=70$ GeV (dotted lines), 
$\E=110$ GeV (dashed lines), 
$\E=149$ GeV (long-dashed lines), 
$\E=243$ GeV (dot-dashed lines).
The horizontal lines mark the corresponding magnitudes of $\chi_s$.}
\label{chi_y.fig}
\end{figure}

Figures \ref{chi_y.fig}(a-b) represent the dependences  $\chi(\rho)$ 
for two temperatures of the crystal and for different energies of the
projectile (as specified). 
For the sake of comparison the values of the quantum parameter
$\chi_s$ (see (\ref{chi_s})) are plotted as well.
The maximum distance $\rho_s$
is defined by the crystal symmetry and the chosen
crystallographic direction and is independent on the crystal temperature.
For a (111) axis in a W crystal $\rho_s=1.292$ \AA. 
As mentioned above the shape of the spectral distribution
(\ref{4.56a}) is quite sensitive to the magnitude of the local 
quantum parameter $\chi(\rho)$. 
The range $\tilde{\rho}_1<\rho<\tilde{\rho}_2$ 
(where $\tilde{\rho}_{1,2}$ are the roots of the equation $\chi(\rho)=1$) 
establishes, for given temperature and energy,
the distances from the axis which contribute  effectively to the emission 
of high energy photons ($\hbar\w\sim \E$) and, consequently, 
define the fraction of spectrum due to the spin-flip transitions.
For fixed temperature the values of $\tilde{\rho}_{1,2}$
do not depend on $\E$, while the 
interval $[\tilde{\rho}_1,\tilde{\rho}_2]$  increases with $T$.
Thus, the intervals of the radial distances which contribute to the 
spin-flip part of the total spectrum ranges from $\approx[0.05,0.18]$ \AA\,
for  $T=70$ K to $\approx[0.0,0.44]$ \AA\, for  $T=293$ K.
Comparing these values with the range $\rho\leq \rho_s=1.292$,
one can anticipate that the radiative spectrum
{\em averaged} over the whole interval of the radial distances 
(see (\ref{4.22})) is  influenced, to a great extent, by the contribution
of the regions where $\chi(\rho)<1$. 
As a result, even for high energies of the projectile electron the relative
yield of soft photons is larger than that of $\hbar\w\sim \E$.

Keeping in mind the large variation in the values of the local quantum
parameter within the interval  $\rho\leq \rho_s$ of the accessible
distances ($\chi(\rho)$ is of the order $10^{-3}$ for 
$\rho \approx \rho_s$, - this range is not presented in figures   
\ref{chi_y.fig}(a-b)), it becomes clear that to obtain realistic 
estimates of the spectral distribution of radiation one has to carry out the
proper averaging procedure.

\begin{figure}
\vspace*{0.cm}
\begin{center}
\epsfig{file=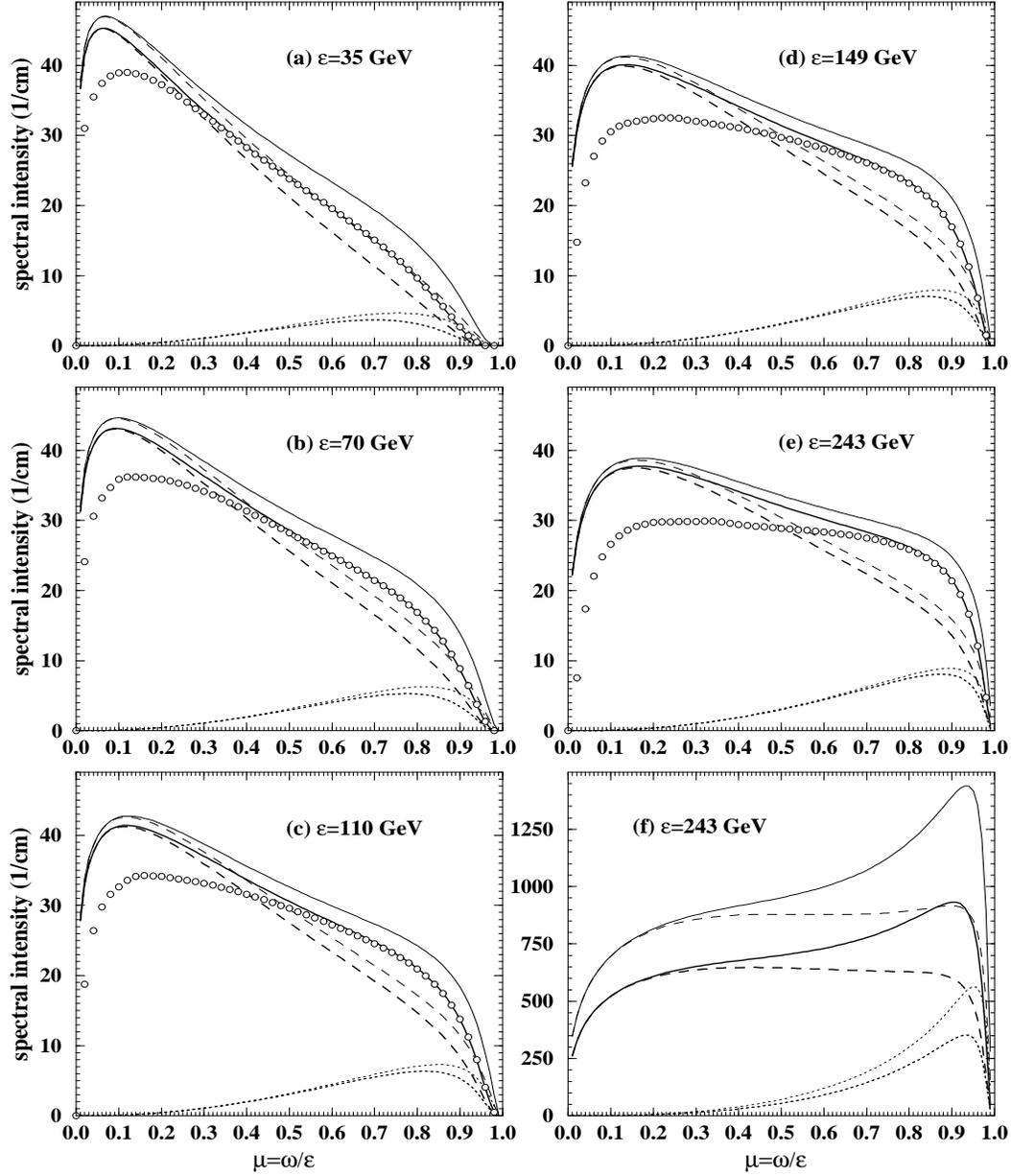, width=14cm, height=16.4cm, angle=0}
\end{center}%
\vspace*{-0.75cm}
\caption{In figures (a-e): the averaged 
spectral distributions of radiation (see (\protect\ref{4.22}))  
calculated for different energies of electrons (as indicated)
aligned 0.3 mrad from the (111) axis in W.
Thick lines refer to the crystal temperature $293$ K, thin lines correspond
to $T=70$ K.
Solid lines and open circles represent the dependences of the total 
intensity,
dotted lines stand for the contribution of the spin-flip transitions,  
dashed lines reflect 
$\langle\overline{{\d E^{(+)}/\hbar\,\d\w\,\d l}}\rangle$.
See also explanations in the text. 
Figure (f) shows the spectral distributions of radiation  from 
a $\E=243$ GeV electron in W at  $T=70$ and $243$ K
calculated without the averaging over the 
$\rho$-interval and for fixed values of the quantum parameter, 
$\chi=\chi_s$.
The legend is as in figures (a-e).}
\label{absolute.fig}
\end{figure}

The results of our calculations of the radiative spectra are presented
in figures \ref{absolute.fig} (a-e). 
The energies of electrons as well as the incident geometry (as specified in 
the caption) correspond to the experimental conditions 
\cite{MikkelsenPRL}.
In addition to the information presented in the caption let us note 
the following. 
The averaging procedure was carried out according to
(\ref{4.22}) where we used $F\left(\rho,\vtheta_0\right)=1$
which is adequate for the over-barrier motion in a thin crystal.
Thus, we have disregarded the evolution of the distribution function
due to the effects of multiple scattering and the radiation damping.
All curves in figures \ref{absolute.fig} (a-e) were obtained within the CFA,
i.e. by using the first two terms in the curly brackets on the right-hand
side of (\ref{4.56a}). 
The dependences represented by open circles were calculated (at $T=293$ K)
using a more elaborated method. For the photon frequencies satisfying the 
condition $l_c \vtheta_0 < a_s$  we used (\ref{4.56a}) including 
the correction terms proportional to ${\vtheta_0^2/ \vtheta_v^2}$.
To calculate the total radiative spectrum in the region of lower photon 
energies we made use of the asymptotic expressions (17.11-12) from
\cite{Baier} which are valid, within logarithmic accuracy, for 
$\chi_s\gg 1$ and $\mu > \chi_s\,(a_s/\rho_s)^{3/2}$.

For the sake of comparison in figure \ref{absolute.fig}(f) we plotted
the spectral distributions  calculated for a $\E=243$ GeV 
electron and for the incident geometry as described above but {\em without}
averaging over the radial distances $\rho$. 
The curves corresponding to $T=70$ and $293$ K were obtained by accounting
for the first two terms in (\ref{4.56a}) where 
the constant values of the quantum parameter $\chi=\chi_s$ (see figure
\ref{chi_y.fig}) were used.
It is exactly this approximation which was used, as one understands, in
\cite{MikkelsenPRL} when interpreting the experimental data. 

First we discuss the differences between the averaged and non-averaged
intensities.
There are three features which clearly distinguish the curves
in figure \ref{absolute.fig} (f) from their analogies in figure 
\ref{absolute.fig} (e).
These are: (a) the absolute magnitude of the intensities,
(b) the shapes of the spectral distributions, and (c) the differences 
between the $T=70$ K and $T=243$ K spectra.

Although striking, these differences can be easily understood.
To illustrate this let us analyze the behaviour of the intensities due
to the spin-flip transitions. 
The analysis of the discrepancies in the total spectral distributions 
can be carried out using similar arguments.

It is easily checked that for large values of $\chi_s$ the corresponding
non-averaged spectral intensity due to the spin-flip transition attains 
its maximum at the photon energies satisfying the condition 
$u = \mu/(1-\mu)\approx \chi_s$.
A quantitative estimate  of the intensity in the vicinity of the maximum reads 
\begin{eqnarray}
\left(\overline{{\d E^{(-)} \over \hbar\,\d \w \, \d l}}
\right)_{\chi=\chi_s\atop u\approx \chi_s}
\approx
-{\alpha \over 3 \lambda_c \gamma}\, \chi_s\, \pAi(1) ,
\label{est.1}
\end{eqnarray}
where $\pAi(1)\approx -0.16$.

To estimate the magnitude of the averaged intensity in the photon energy range
$u \approx \chi_s$ one first notices that the principal contribution comes from
the spatial region where the local quantum parameter $\chi(\rho)$ is greater 
than $\chi_s$ (see figure \ref{chi_y.fig}). Denoting 
 the roots of the equation $\chi(\rho)=\chi_s$ through $\rho_{1,2}$, and
taking into account that $\rho_{1}^2\ll \rho_{2}^2$ one obtains the 
following estimate:
\begin{eqnarray}
\fl
\left\langle\overline{{\d E^{(-)}\over\hbar\d\w\,\d l}}
\right\rangle_{u\approx\chi_s}
\approx
-{\alpha \over 3 \lambda_c \gamma}\,\chi_s 
\int_{\rho_1}^{\rho_2}{2 \rho \d\rho \over \rho_s^2}\,
{\pAi(z) \over z}
\approx
-{\alpha \over 3 \lambda_c \gamma}\,
\eta\,
\langle\chi\rangle\,
\langle z\rangle^{1/2}\pAi\left(\langle z\rangle\right),
\label{est.2}
\end{eqnarray}
where $\langle\chi\rangle$ is the average value of the quantum 
parameter over the region $\rho_1 \leq \rho \leq \rho_2$.
For the temperatures 70 and 293 K $\langle\chi\rangle$ equals
to $21.1$ and $12.3$, respectively.  
The quantity  $\langle z\rangle$, defined as 
$\langle z\rangle = (\chi_s/\langle\chi\rangle)^{2/3}$,
equals  $0.65$ for $T=70$ K and 
$0.71$ for $T=293$ K.
One easily checks (see, e.g.,\cite{Abramowitz}) that for both of these
values 
$\langle z\rangle^{1/2}\pAi\left(\langle z\rangle\right)\approx -0.167$.
The parameter $\eta$ stands for the ratio ${\rho_2^2/ \rho_s^2}$ yielding
the values  $\eta \approx 0.012$ for $T=70$ K and 
$\eta \approx 0.019$ for $T=293$ K.
The factor $\chi_{eff}\equiv \eta\,\langle\chi\rangle$ 
represents by itself the effective
quantum parameter which defines the absolute magnitude of the averaged 
intensity in the range of the photon frequencies $\hbar\w \sim \E$.

The right-hand sides of eqs. (\ref{est.1}) and (\ref{est.2}),
reproducing with a reasonable accuracy 
the maximum values of the spectral intensities due to the spin-flip 
transition (the dotted curves in figures \ref{absolute.fig} (e) and (f)),
allows one to explain the discrepancies between the averaged and non-averaged
spectra.
The excess of the non-averaged intensities over the averaged ones is reproduced
by the ratio $\chi_s/\chi_{eff}$ which equals $44$ at $T=70$ K and 
$31$ at $T=293$ K which are close to the exact values which follow from
figures \ref{absolute.fig} (e) and (f): $57$ and $43$, respectively. 
The different behaviour of the two types of the spectral distributions
also follows from the above written formulae.
Indeed, the ratio $\chi_s\left(T=70\right)/\chi_s\left(T=293\right)=1.7$ 
explains the difference between the thick and thin dotted lines in 
figures \ref{absolute.fig} (e), whereas the ratio
$\chi_{eff}\left(T=70\right)/\chi_{eff}\left(T=293\right)\approx 1.1$
illustrates that the averaged intensities are much less sensitive to the
change in the crystal temperature.

The behaviour of 
$\langle\overline{{\d E^{({\rm tot})}/\hbar\,\d\w\,\d l}}\rangle$, 
represented by solid lines in figures \ref{absolute.fig} (a-e),
illustrates the known tendencies of the total spectral distributions 
obtained within the CFA \cite{Baier}.
Namely, for comparatively low incident energies, when the local quantum
parameter  is small in a wide interval of the distances
(see figure \ref{chi_y.fig}) the spectrum has maximum at $\hbar\w \ll \E$
and decreases rapidly with the photon energy.
As $\E$ increases, the range where $\chi(\rho)\gg 1$ increases
as well giving rise to the yield of the photons with $\hbar\w \sim \E$.
The magnitude of the intensity in the low-$\w$ region decreases and the spectrum
becomes flatter over the whole range of photon energies.
 
Similar features  characterize the total spectral distributions obtained 
from (\ref{4.22}) and (\ref{4.56a}) 
including the correction terms proportional to 
${\vtheta_0^2/ \vtheta_v^2}\approx 0.16$, - the open circles in figures
\ref{absolute.fig} (a-e).
The contribution of the correction terms is negligibly small for
$\hbar\w \sim \E$ when the coherence length is small enough so that the
characteristics of the spectrum are defined by the instant
value of the particle's 
acceleration. 
Their influence becomes visible in the low-energy part of the spectrum. 
As it is seen, the inclusion of the  correction terms 
decreases the magnitude of the spectrum for $\hbar\w \ll \E$ which leads to 
a further flattening of the distribution.

Commenting on the spectral intensities due to the spin-flip transitions
we note the following.
For all incident energies the spectra 
become pronounced close to the tip-region. 
This feature, which is known for both the synchrotron radiation
\cite{Land4,Sokolov,Baier1973} and for the radiation by electrons in 
the field of crystals \cite{MikkelsenPRL,Baier,Lindhard91,Sorensen1996},
can be easily traced from the general formulae (\ref{4.22a}-\ref{4.22d}).
Indeed, the spin-flip intensity is proportional to the factor
$\mu^3/(1-\mu)$ which is small for $\mu\ll 1$ and increases as $\mu\sim 1$. 
For $\mu\longrightarrow 1$ the spectrum 
$\overline{{\d E^{(-)}/\hbar\,\d\w\,\d l}}$ exponentially decreases
due to the highly oscillatory exponential factor in the integrand in 
(\ref{4.22a}).
The contribution to the averaged spectrum (\ref{4.22}) 
comes from the spatial region where $\chi(\rho)\gg 1$. 
The position of the maximum of 
$\langle\overline{{\d E^{(-)}/\hbar\,\d\w\,\d l}}\rangle$  can be estimated 
as $\mu \approx 1 - \langle \chi \rangle^{-1}$, see (\ref{est.2}). 
The latter formula, as it is mentioned above, allows to estimate the 
magnitude of the maximum intensity, which is almost 
independent of $\gamma$ for sufficiently high energy of the electron. 
Indeed, apart from the explicit factor in the denominator,
 the dependence on $\gamma$ enters the right-hand side of (\ref{est.2})
through  $\langle \chi \rangle$ and  $\langle z \rangle$.
Since the ratio $\langle \chi \rangle / \gamma $ is independent of the 
relativistic factor (see (\ref{chi_s}) and (\ref{4.61b})), the dependence
in question is concentrated in 
$\langle z\rangle^{1/2}\pAi\left(\langle z\rangle\right)$
which is a slowly increasing function of $\gamma$. 
This feature is clearly seen in the figures: the maximum of 
$\langle\overline{{\d E^{(-)}/\hbar\,\d\w\,\d l}}\rangle$ increases
by a factor of 2 with the energy variation from $35$ to $243$ GeV.  

Let us mention that the spin-flip spectral distributions were 
obtained within the CFA, omitting the correction terms on the right-hand 
side of  (\ref{4.56a}).
These terms contribute noticeably to the region $\mu \ll 1$ but are negligibly
small closer to the hard-end of the spectrum.

Finally, let us present the numerical data on the  
effective radiation length, $L_{eff}$. 
This quantity, which  is defined as
\begin{eqnarray}
L_{eff}^{-1}=
\int_0^{\E}
\left\langle
\overline{{\d E^{({\rm tot})} \over \d \w \, \d l}} 
\right\rangle\, {\d\w \over \E},
\label{Leff}
\end{eqnarray}
defines the interval within which a particle loses
its energy due to the radiation.

The values of $L_{eff}^{-1}$, calculated using the
averaged CFA spectra with the correction terms included (open circles in the
figures), and the ratios 
$r=L_{eff}^{-1}/L_{BH}^{-1}$, where 
$L_{BH}^{-1}\approx 4Z^2\alpha r_0^2\, n \, \ln\left(183Z^{-1/3}\right)$ 
($n$ is the volume density of the crystal atoms, $r_0$ is the electron 
classical radius, $Z$ is the atomic number) 
is the radiation length due to bremsstrahlung in 
an amorphous medium, are summarized in table \ref{Table}.
The values of $r$ are in good agreement with the measured data
\cite{MikkelsenPRL} except for the case $\E=35$ GeV where our result
is $\approx 10$\% higher.
\begin{table}
\caption{The values of $L_{eff}^{-1}$ (see (\protect\ref{Leff}))
and $r=L_{eff}^{-1}/L_{BH}^{-1}$ 
calculated for different energies of electron.}
\begin{indented}
\item[]\begin{tabular}{@{}rrrrrr}
\br
$\E$ (GeV)                & 35  & 70  & 110 & 149 &  243    \\
$L_{eff}^{-1}$ (cm$^{-1})$&21.50&23.89&25.64&25.77&25.86  \\
$r$                       & 7.10& 7.88& 8.46& 8.51& 8.54 \\
\br
\end{tabular}
\end{indented}
\label{Table}
\end{table}

\begin{figure}
\vspace*{0.cm}
\begin{center}
\epsfig{file=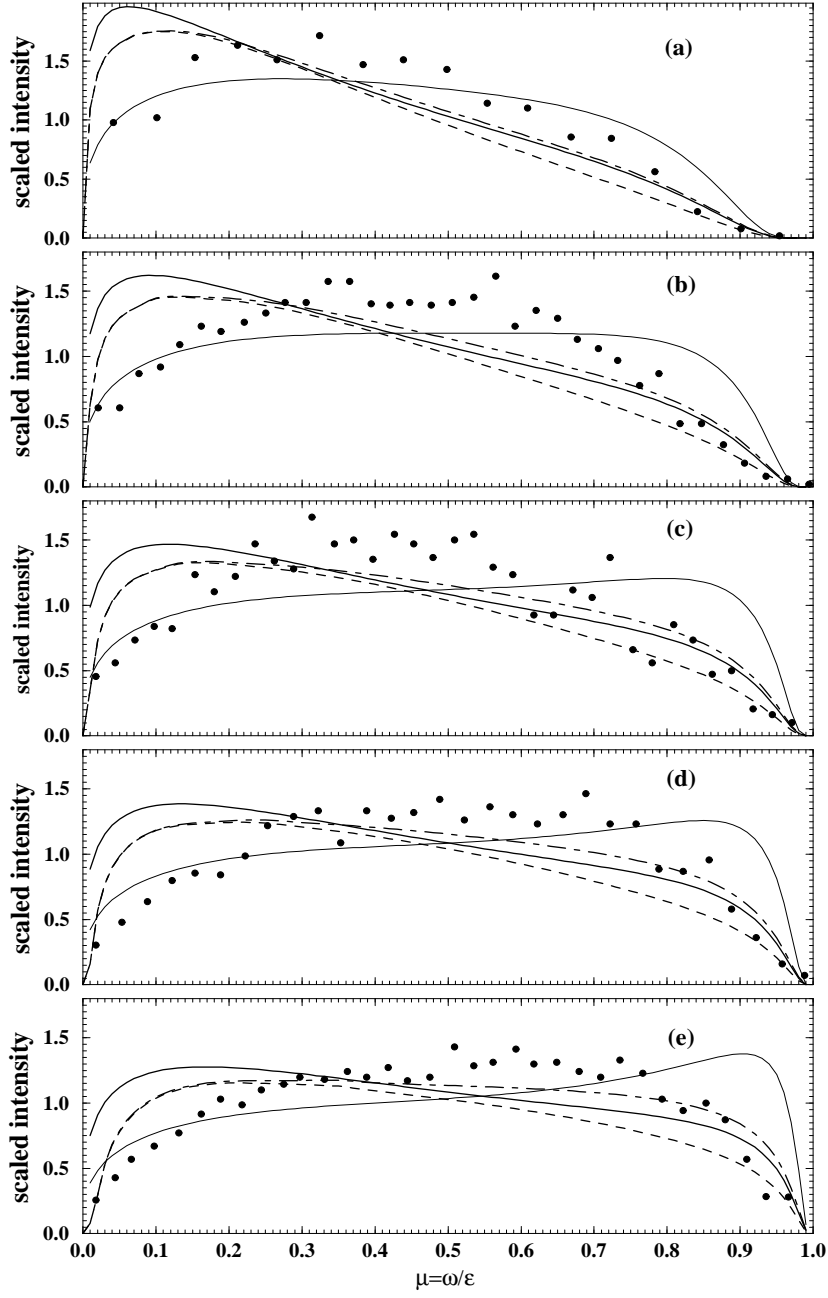, height=17.3cm,width=11cm,  angle=0}
\end{center}%
\vspace*{-0.75cm}
\caption{
Scaled  intensities  for
different incident energies: 
(a) $\E=35$ GeV, (b) $\E=70$ GeV,
(c) $\E=110$ GeV, (d) $\E=149$ GeV,
(e) $\E=243$ GeV.
Thick solid lines correspond to the total intensities  within CFA.
Thick dot-dashed lines represent the total intensities 
and thick dashed lines stand for the intensities 
without spin-flip  both 
calculated within the approximation 'CFA + correction terms'.
Thin solid lines denote the distributions 
calculated without the averaging over the 
$\rho$-interval and for fixed values of the quantum parameter, 
$\chi=\chi_s$.
All curves correspond to $T=293$ K.
Full circles are the experimental data \protect\cite{MikkelsenPRL}.}
\label{experiment.fig}
\end{figure}

In figure \ref{experiment.fig} we present the results of our calculation 
together with the experimentally measured spectra \cite{MikkelsenPRL}.
The data  refer to the scaled spectral intensity, i.e. 
$\langle\overline{{\d E^{({\rm t})}/\hbar\,\d\w\,\d l}}\rangle$ 
divided by $L_{eff}^{-1}$.
The crystal temperature is chosen to be $293$ K. 
It is this $T$-value which had been used in the experiment \cite{MikkelsenPRL} 
as one can deduce from the paper basing on the cited values of the 
parameter $\chi_s$. 
Let us note that we failed to reproduce the curves presented in the cited 
paper and which, as it follows from the text, correspond to the CFA 
approximation with constant values $\chi=\chi_s$ 
without averaging over the distances $\rho$ (full lines in figure 1
in \cite{MikkelsenPRL}). 
In figure \ref{experiment.fig} thin lines correspond to the calculation
of 
$\overline{{\d E^{({\rm tot})}/\hbar\,\d\w\,\d l}}$ 
according to (\ref{4.56a}) but omitting the terms proportional
to ${\vtheta_0^2 / \vtheta_v^2}$ and using the values $\chi_s$ (as indicated
in figure \ref{chi_y.fig}(b)) instead of the local parameter 
$\chi(\rho)$. 
Although it seems like the same approximation is used the shapes of 
these curves noticeably differs from those presented in the cited paper.
In any case this approximation, in our mind, is not 
adequate to describe the radiative spectra in the crystal (see the discussion
in connection with figure \ref{absolute.fig}), so that the 
thin solid curves are plotted for the illustrative purposes only.

'The scattering of the experimental points indicate the uncertainties'
\cite{MikkelsenPRL}.
Taking this into account we may state
that for $\mu > 0.7$ both thick curves, which describe the total 
intensities, match the experimental data.
Note that it is exactly the region where the spin-flip transitions contribute
to the total spectrum (see  figure \ref{absolute.fig}).
In the mid-range of the $\mu$-values the computed intensities underestimate
the experimental ones, and for $\mu \ll 1$ theoretical results noticeably 
overestimate the measured spectra.
All these features are typical, as it is seen from the figures, for 
the whole range of the incident energies, and, in general, the calculated 
curves are skewed to the left wing of the spectrum as compared to the 
measured data.

For the sake of comparison we plotted also the scaled intensities 
of the radiation excluding the contribution of the spin-flip transitions
(dashed lines). 
These were obtained as 
$L_{eff}
\left(\langle\overline{{\d E^{({\rm tot})}/\hbar\,\d\w\,\d l}}\rangle
- \langle\overline{{\d E^{(-)}/\hbar\,\d\w\,\d l}}\rangle\right)$ 
with both terms in the brackets calculated with the help of 
(\ref{4.22}) and (\ref{4.56a}), and retaining the correction terms in the
latter formula.
Taking into account the experimental uncertainties one can observe that 
these curves taken alone agree fairly well with the measured data
on the right wing of the spectrum.

Thus, on the basis of these calculations we cannot confirm in full
the statement made in \cite{MikkelsenPRL} that the region $\mu\sim 1$
of the radiative spectrum by ultra-relativistic electrons moving under a
small incident angle with respect to the crystallographic axis is formed
mainly due to the spin-flip transition of the projectile.
 
\section{Concluding remarks\label{Conclusion}}

We have calculated the spectral distributions of radiation
formed during the passage of multi-GeV unpolarized electrons 
along (111) axis in W crystal.
The consideration was carried out  within the framework of the quasi-classical
approach \cite{Baier} which adequately describes the radiative process of
ultra-relativistic particles in strong external fields.

Special attention was paid to the investigation of the contribution of the 
spin-flip transitions of the projectile to the total spectrum. 
The results of our study can be summarized as follows.
The role of the spin-flip transitions in forming the hard-photon end,
$\hbar\w \approx \E$, of the radiative spectrum is quite noticeable.
Definitely, it is much larger than that predicted in \cite{Augustin}
where a full quantum-electrodynamical approach was developed.
However, we also can hardly agree with the statement made 
in \cite{MikkelsenPRL}
that it is the spin-flip transitions which do modify dramatically the 
shape of spectral distributions.
The numerical results presented in this paper agree well with the experimental
data in the range of photon energies $\hbar\w \approx (0.7\dots 1.0)\E$, i.e.
exactly where the contribution of the spin-flip transitions becomes pronounced.
This fact demonstrates the applicability of the standard CFA model (combined
with a proper procedure for averaging over the phase volume of a projectile)
for the description of the hard-end of the spectrum for over-barrier particles.
For lower photon energies, when the coherence length becomes larger than the
typical scale of variation of the potential of a crystallographic axis/plane,
it is necessary to go beyond the CFA scheme. 
In the present paper we utilized the approach suggested in \cite{Baier}
which allows to incorporate the corrections proportional to 
$(\vtheta_0/\vtheta_v)^2$. 
Our data shows that even when this parameter is small (in the case considered
in this paper $(\vtheta_0/\vtheta_v)^2=0.16$) the correction terms 
reduce noticeably the magnitude of the spectral intensity in the range of 
small photon energies.
This approach becomes less adequate as $\hbar\w/\E \ll 1$. 
In this case one has to use the formulae (\ref{4.22a}-\ref{Delta.2})
which include the deviations from the CFA in the most general way.

The theoretical results can be improved further by taking into account
three important effects which are intrinsic for the radiative scattering of
an ultra-relativistic particle in the field of a crystal.
The first one is the analogue of the Landau-Pomeranchuk-Migdal effect
\cite{LPM}. 
The latter consists in a strong suppression of the yield of the soft-photon
bremsstrahlung emitted by an ultra-relativistic particle in medium.
This is a result of a number of correlated small-angle collisions of the
projectile with the lattice atoms on the scale of the coherence length.
Similar effect occurs when the particle moves close to an atomic string
in a crystal and under the small angle $\vtheta_0$ from the axis \cite{Fomin}.
The account for this mechanism will reduce the magnitude of the 
spectral intensity for low photon energies.
Another phenomenon to be accounted for is the cascade process 
of the photon emission which is effectively generated when 
a multi-GeV electron/positron
enters the crystal at small ($\vtheta_0<\vtheta_v$) angles to the crystallographic
axis/plane \cite{Baier,25pdf}.
The third effect we would like to mention is the influence of the radiative loss on 
the shape of the emission spectrum.
In the present calculations we disregarded 
the change in the particle's energy on the scale of the crystal thickness 
$L$.
However, as soon as $L$ becomes comparable with the effective 
radiation length the formalism must include the solution of the proper
kinetic equation (\cite{Baier,Augustin}). 
This consideration is fully appropriate in the case considered in the present
paper with regard to the experiment  \cite{MikkelsenPRL}.
Indeed, the experimental setup included the $0.2$ mm thick W crystal.
Comparing this length with the values of $L_{eff}$ (see table \ref{Table})
one finds  $L/L_{eff}\approx 0.5$, so that the electron looses half of its
initial energy. 
Although it is clear, qualitatively, how the radiative loss 
influences the radiative spectrum (the latter is shifted towards the 
low-energy part), the rigorous quantitative treatment is to be
performed for each specific set of the parameters: the incident energy and 
the geometry, and the type of the crystal.

Finally, let us mention the phenomenon which was left beyond the scope of this paper
 and which, as far as we know, has not been studied neither theoretically nor
experimentally.
It is well-known (e.g.  \cite{Sokolov}) that the magnetic bremsstrahlung leads
to the radiative self-polarization of the electron. 
Formally, this is due to the fact that the differential probability contains the
term proportional to $\bxi \cdot{\bf H}$ (here ${\bf H}$
is the vector of the external
magnetic field) which depends on the angle between $\bxi$ and ${\bf H}$.
A similar effect was studied in \cite{Bagrov} in connection with the axial 
channeling of electrons/positrons.
If one takes into account the radiative loss (which is the case when $\chi> 1$
and $\gamma \gg 1$) then there appears a strong interaction of the spin with 
the field of radiation, ${\bf H}_r$ \cite{Baier1973}. 
This results in (a) a noticeable change of the typical time $\tau$ of 
the radiative self-polarization, and (b) in radiative damping which leads to the decrease
of $|\bxi|$.
Later these phenomena were considered in connection with the planar 
channeling
of ultra-relativistic electrons/positrons in slightly bent crystals
\cite{Barysh79,Barysh82,Mikhalev}.
It was demonstrated that there appear additional terms, due to the 
centrifugal force in the bent channel, which influence the spin-precession, and, as
a result, the time of the self-polarization.
We note here that the planar channeling is more preferable than the axial one
if studying the effect of radiative self-polarization. 
The reason is that the terms proportional to $\bxi{\bf H}$ (${\bf H}$ is the effective
magnetic field acting on the particle in the channel) are less influenced by the
averaging over the phase distribution in the planar case.

In our mind it will be very interesting to carry out the analysis
(theoretically and, if 
feasible, experimentally as well)
of the self-polarization of a bunch of ultra-relativistic particles undergoing
channeling in a crystal whose channels are periodically bent. 
In this case there arises very intensive radiation, additional to the well-known 
channeling radiation, due to the periodicity of the trajectory of the 
projectile \cite{We}.

\ack

The authors are grateful to Prof. E.~Uggerh{\o}j and Dr.
U.~Mikkelsen for sending the manuscript of the paper \cite{MikkelsenPRL}
prior to its publication.

The research was supported by DFG, GSI, and BMFT.
AVK and AVS acknowledge the support from the 
Alexander von Humboldt Foundation.


\end{document}